\documentclass[smallextended]{svjour3}       
\smartqed  
\usepackage{graphicx}
\usepackage{amsmath}
\usepackage{amssymb}

\newcommand{\be}{\begin{equation}}
\newcommand{\ee}{\end{equation}}
\newcommand{\bea}{\begin{eqnarray}}
\newcommand{\eea}{\end{eqnarray}}

\newcommand{\refc}[1]{(\ref{#1})}
\newcommand{\Ord}{O}

\newcommand{\bt}{\bar{b}_2}
\newcommand{\btt}{$\bt$}

\newcommand{\SU}{\text{SU}}

\begin{document}

\title{Revisiting the flux tube spectrum of 3d $\SU(2)$ lattice gauge theory}
\author{Bastian B. Brandt}
\institute{Fakult\"at f\"ur Physik, Universit\"at Bielefeld,
D-33615 Bielefeld, Germany\\
\email{brandt@physik.uni-bielefeld.de}}

\date{Received: date / Accepted: date}

\maketitle


\abstract{
We perform a high precision measurement of the spectrum of the flux tube
in three-dimensional $\SU(2)$ gauge theory at multiple lattice spacings.
We compare the results at large $q\bar{q}$ separations $R$ to the spectrum
predicted by the effective string theory, including the leading order boundary
term with a non-universal coefficient. We find
qualitative agreement with the predictions from the leading order Nambu-Goto
string theory down to small values of $R$, while, at the same time, observing
the predicted splitting of the second excited state due to the boundary term.
On fine lattices and at large $R$ we observe slight deviations from the EST
predictions for the first excited state.

}

\section{Introduction}

While the microscopic origin of confinement in Quantum Chromodynamics (QCD)
remains elusive due to the lack of analytic methods to solve QCD at small energies,
the formation of a region of strong chromomagnetic flux, a flux tube, between quark
and antiquark provides a heuristic explanation for quark confinement. Strong evidence
for flux tube formation has been found in numerous simulations of lattice QCD, both
in the quenched approximation, e.g.~\cite{Bali:1994de}, and in simulations with dynamical
fermions, e.g.~\cite{Cea:2017ocq} (for reviews and more detailed lists of references
see~\cite{Bali:2000gf,Brandt:2016xsp}).

For large quark-antiquark distances $R$ the flux tube is expected to resemble a thin
energy string, so that its dynamics will be governed by an effective (bosonic) string
theory (EST). Here and in the following we neglect the effects of dynamical quarks,
which enable the formation of a light quark-antiquark pair from the vacuum and a
breaking of the string. Following two decades of tremendous progress, a number of
features of the EST, including the spectrum up to
$O(R^{-5})$~\cite{Aharony:2010db,Aharony:2011ga,Caselle:2013dra},%
\footnote{The EST is formulated as a derivative expansion of the Goldstone bosons
associated with the breaking of translational symmetry by the presence of the string,
effectively leading to an expansion of the string related observables in $R^{-1}$.}
are by now rather well understood. See Refs.~\cite{Aharony:2013ipa,Brandt:2016xsp}
for recent reviews.
\footnote{Note, that there is still a discrepancy in the literature between the
results from Aharony and collaborators on effective string theories and the results
on conformal string theories by Hari Dass (and initially Drummond) and collaborators
\cite{Drummond:2004yp,HariDass:2006sd,HariDass:2007dpl,HariDass:2009ub,HariDass:2009ud,Dass:2009xe}.
See also the contribution of Hari Dass to this memorial issue for more details.}
Most of the parameters in the EST are constrained by Lorentz
symmetry and take universal values. The first non-universal parameter in the EST,
denoted as \btt{} (in its dimensionless version), appears at order $\Ord(R^{-4})$
and is the coefficient of a boundary term.

The spectrum of the flux tube excitations can be computed in numerical simulations
of pure gauge theories, where dynamical fermions are absent.
Typically good agreement between the EST predictions and the lattice results has
been observed down to $q\bar{q}$ separations where the EST is expected to 
break down (for a compilation of results see~\cite{Brandt:2016xsp}). Since the
non-universal coefficients appear in subleading terms, their extraction requires
very accurate results for the energy levels up to large values of $R$. So far
sufficient accuracy has only been achieved for the coefficient
\btt{} in three dimensions
(3d)~\cite{Brandt:2010bw,Billo:2012da,Brandt:2013eua,Brandt:2017yzw,Brandt:2018fft},
where the simulations are at least an order of magnitude less expensive than in the
4d case. First hints for a non-vanishing \btt{} in 4d $\SU(3)$ gauge theory, however,
have recently been obtained both from the groundstate energy -- the static potential --
and the flux tube profile at non-zero temperature~\cite{Bakry:2018kpn,Bakry:2020ebo}.
Despite a parametric suppression, the most precise computation of \btt{} comes from
the static potential, which can be obtained with much higher accuracy than the
excitation spectrum. The consistency between the boundary coefficient extracted from
the static potential~\cite{Brandt:2017yzw} and the excited states has so far only been
checked for a comparably coarse lattice
spacing~\cite{Brandt:2010bw,Brandt:2018fft}. In this article, we will extend and improve
on our initial studies of the excitation spectrum in 3d $\SU(2)$ gauge theory,
Refs~\cite{Majumdar:2002mr,Brandt:2007iw,Brandt:2009tc,Brandt:2010bw}.
In particular, we aim at investigating the continuum approach of the excited states
at large $R$, which has not been possible with sufficient accuracy previously, and compare
our results to the EST predictions using the EST parameters extracted
in Ref.~\cite{Brandt:2017yzw}.

Currently, the main challenge from the theory side concerns the
inclusion of corrections due to the vorticity of the flux tube, possibly showing
up as massive modes on the worldsheet
(e.g.~\cite{Polyakov:1986cs,Klassen:1990dx,Nesterenko:1997ku,Caselle:2014eka};
see also the more detailed discussion and list of references in
Ref.~\cite{Brandt:2017yzw}).%
Candidate states with contributions from massive modes have been seen in 4d
$\SU(N)$ gauge
theories~\cite{Morningstar:1998da,Juge:2002br,Juge:2004xr,Athenodorou:2010cs}.
For closed strings the results for such an anomalous state are in good
agreement with the contribution of a massive pseudoscalar particle on
the worldsheet, known as the worldsheet
axion~\cite{Dubovsky:2013gi,Dubovsky:2014fma} (see also~\cite{Bakry:2020akg}).
Similar anomalous states do not appear in 3d, which might be related to
the absence of the topological coupling term and, consequently, less
sensitivity of the energy levels to massive modes. In 3d, the presence of
massive modes leads to an additional term which mixes with the boundary
correction term at $\Ord(R^{-4})$ and, consequently, impacts the numerical
result for \btt{}. Since the 3d massive mode contributions to the excited
state energies are unknown, we cannot perform a direct comparison of the
cases with and without massive modes as done for the static potential
in~\cite{Brandt:2017yzw} and leave this for future studies.

\section{Excited states of the QCD flux tube}
\label{sec:setup}

In this section we will focus on the extraction of the excited states
of the flux tube in 3d $\SU(2)$ gauge theory for multiple lattice spacings.
The comparison with the EST is left for the next section.

\subsection{Computation of Wilson loops for excited states}

While for the extraction of the static potential the ideal observables
are Polyakov loop correlation functions
(e.g.~\cite{Luscher:2002qv,HariDass:2007tx,Brandt:2017yzw}), spatio-temporal
Wilson loops of extents $R$ and $T$, including operators coupling to particular
quantum number channels on the spatial lines, are suitable observables for the
extraction of excited states. The main difficulty is the need for large loops
for which the signal-to-noise ratio decreases exponentially with the area of
the loops. At a given value of the $q\bar{q}$ distance $R$, the reliable
high-precision extraction of the energy state demands the sufficient
suppression and control of the contributions of excited states to the Wilson
loop expectation value (cf. eq.~\refc{eq:ev-spectral}). In the first comprehensive
study, large correlation matrices including highly optimized operators
in combination with smearing and anisotropic lattices have been used to
extract the excitation spectrum in different gauge theories and three and four
dimensions~\cite{Juge:2002br,Juge:2003ge,Juge:2004xr}. After careful diagonalisation,
excited state contributions are suppressed in the individual eigenvalue correlators,
~\footnote{Using a generalised eigenvalue problem in a suitable setup, excited state
contaminations only include states starting from the $N$-th excited state in the
particular quantum number channel, where $N$ is the number of operators included in
the correlation matrix (see Ref.~\cite{Blossier:2009kd}).}
so that the convergence to the $T\to\infty$ limit is enhanced for the low-lying
states. An alternative
strategy has been pursued in~\cite{Majumdar:2002mr,Brandt:2007iw,Brandt:2009tc},
using an improved version of the L\"uscher-Weisz multilevel
algorithm~\cite{Luscher:2001up} for error reduction to reliably extract large
loops for a small number of spatial operators and temporal
extents. The residual excited state contaminations have been removed using
suitable fit functions.

Both strategies give reliable results for intermediate values of $R$, but for
large values of $R$ the contamination due to excited states becomes more and more
severe, since the energy gaps to the excited states decrease.
Consequently, there is doubt about the sufficient suppression of the excited
contributions in this regime. A combination of the two methods, using the improved
multilevel algorithm in combination with a large set of operators, correlation matrices
and fits including excited state contributions, has been used in
Ref.~\cite{Brandt:2010bw}, leading to accurate results for the excited states in 3d
$\SU(2)$ gauge theory at large values of $R$, albeit at a single and comparably
large lattice spacing. In this study we use this strategy, further improving on the
analysis by employing improved fits and more temporal extents in the analysis, to
extract the spectrum at smaller lattice spacings.

In 3d the string energy levels can be classified by the quantum numbers
of charge conjugation and parity $(C,P)$. The individual combinations are denoted as
channels. Using suitable sets of spatial operators $S_i^j$, the correlation
matrices with respect to these quantum numbers can be transformed to a block-diagonal
form using the linear combinations
\begin{equation}
\label{eigstates}
\begin{array}{l}
S_{i}^{++} = S_{i}^{1} + S_{i}^{2} + S_{i}^{3} + S_{i}^{4} \\
S_{i}^{+-} = S_{i}^{1} + S_{i}^{2} - S_{i}^{3} - S_{i}^{4} \\
S_{i}^{--} = S_{i}^{1} - S_{i}^{2} - S_{i}^{3} + S_{i}^{4} \\
S_{i}^{-+} = S_{i}^{1} - S_{i}^{2} + S_{i}^{3} - S_{i}^{4} \; .
\end{array}
\end{equation}
To extract excited states in a given $(C,P)$ channel, we use the 8 different operator
sets introduced in Ref.~\cite{Brandt:2010bw}. Setting up a generalized eigenvalue problem
for these correlation matrices for the reliable extraction of the eigenvalues in the limit
$T\to\infty$ is problematic, since the correlation matrices might be ill-conditioned
already for the smallest temporal extent available. As in Ref.~\cite{Brandt:2010bw}, we
thus diagonalize the correlation matrices for each value of $T$ separately utilizing
the $QR$ reduction method. The resulting eigenvalues are denoted as $\lambda^{CP}_{n}(R,T)$
with $n=0,\,\ldots,\,7$. To check that we are identifying the right eigenvalues with
increasing $T$, we look at the associated eigenvectors. Those belonging to the same states
for different $T$ should be almost parallel, whereas those belonging to different states
perpendicular. In the following we will mostly focus on the groundstates in the
individual channels, for which the identification of the eigenvalues is unambiguous.
The only excited state which we consider is the first excited state in the $(+,+)$-channel,
for which the identification becomes more difficult at smaller lattice spacings.
In contrast to the eigenvalues obtained from a generalized eigenvalue problem, the resulting
eigenvalues might include contaminations from other states in the channel, albeit with
overlaps which are strongly suppressed for $T\to\infty$.

\subsection{Removal of excited state contaminations}
\label{sec:exc-state-fits}

The eigenvalues of the correlation matrices generically obey the spectral
representation~\cite{Luscher:1990ck}
\begin{equation}
\label{eq:ev-spectral}
 \lambda^{CP}_{n}(R,T) = \beta^{CP}_n(R) e^{-E^{CP}_{n}(R)\:T}
 \Big[ 1 + \sum_{k=0\neq n}^{\infty} \widetilde{\alpha}_{k,n}^{CP}(R) \: e^{-|\Delta E^{CP,CP}_{kn}(R)|\:T} \Big] .
\end{equation}
Here $E^{CP}_{n}$ are the energies in the $(C,P)$-channel,
\begin{equation}
\label{eq:en-diffs}
\Delta E^{CP;CP'}_{nm}(R) \equiv E^{CP}_n(R)-E^{CP'}_m(R)
\end{equation}
are the energy differences, $\beta^{CP}_n$ the overlap of the eigenstate with energy state $n$
and $\displaystyle \widetilde{\alpha}_{k,n}^{CP}(R)$ the ratio of overlaps of the eigenstate with
energy states $k$ and $n$. At finite $T$, it depends on the size of the energy gaps and the reduction
of overlaps which state dominantes the sum on the right hand side of eq.~\refc{eq:ev-spectral}.

At large $T$ and with sufficient suppression of contributions from other states, the energies can be
extracted using the asymptotic $T\to\infty$ formula
\begin{equation}
\label{eqasympt}
-\ln\left(\lambda^{CP}_{n}(R,T)\right) = \bar{E}^{CP}_{n}(R)\:T - \ln\left(\beta^{CP}_{n}(R)\right) \; .
\end{equation}
In practice, however, contaminations from other states are not negligible due to the high accuracy
for the eigenvalues, so that eq.~\refc{eqasympt} cannot be used to extract the energies reliably.
To remove contaminations from other states we perform a simultaneous fit to the results for
the eigenvalues for all available $T_a$ and $T_b$ to the leading order formula
(see also \cite{Majumdar:2002mr,Brandt:2009tc,Brandt:2010bw})
\begin{equation}
\label{eqfit1}
\begin{array}{rl}
\displaystyle  - \frac{1}{T_{b}-T_{a}} \: \ln \left[ \frac { \lambda^{CP}_{n} ( R , T_{b} ) }
{ \lambda^{CP}_{n} ( R , T_{a} ) } \right] = & \displaystyle E^{CP}_{n} ( R ) +
\frac{1}{T_{b}-T_{a}} \: \alpha_n^{CP}(R) \: e^{ - \delta_n^{CP}(R) \: T_{a} } \vspace*{2mm} \\
 & \displaystyle \times \left( 1 - e^{ - \delta_n^{CP}(R) \: ( T_{b} - T_{a} ) } \right) \: .
\end{array}
\end{equation}
Here $T_a<T_b$, $\displaystyle \alpha_n^{CP}(R)\equiv \widetilde{\alpha}_{n+1,n}^{CP}(R)$ and
$\delta_n^{CP}(R)$ is the energy gap to the closest state (or the one with the dominant
contribution to the sum on the right-hand-side of eq.~\refc{eq:ev-spectral}) in the channel.
In most of the cases these fits lead to accurate results with acceptable control over the systematic
effects. In some cases, however, these fits become unstable with respect to statistical fluctuations
of the eigenvalues. This is particularly true for large values of $R$ and higher excited states.
To further constrain the fits, we include the data for the individual eigenvalues in the global
fit, leading to the additional relations
\begin{equation}
\label{eqfit2}
- \frac{1}{T_{b}} \: \ln \left[ \lambda^{CP}_{n} ( R , T_{b} ) \right] =
 E^{CP}_{n} ( R ) - \frac{1}{T_{b}} \: \left( \gamma_n^{CP}(R) +
\alpha_n^{CP}(R) \: e^{ - \delta_n^{CP}(R) \: T_{b} } \right) \: ,
\end{equation}
where $\gamma_n^{CP}(R)=\ln\big(\beta^{CP}_{n}(R)\big)$ is an additional fit parameter. Despite
this additional parameter, the joint fits using eqs.~\refc{eqfit1} and~\refc{eqfit2} are
generically more stable. A particular example for such a fit is shown in Fig.~\ref{fig:frem-fit} in
appendix~\ref{app:exc-fits}. To control the systematics in these fits we implement further checks
which are also described in appendix~\ref{app:exc-fits}.

The energy differences $\Delta E^{CP;CP'}_{nm}(R)$ from eq.~\refc{eq:en-diffs}
can be extracted independently from the total energies, so that they
might serve as independent crosschecks.
Using Eq.~\refc{eq:ev-spectral}, one can derive the analogue to
eq.~\refc{eqfit1} for the energy differences
\begin{equation}
\label{eqfitdiff1}
\begin{array}{rl}
- & \displaystyle \frac{1}{T_{b}-T_{a}} \: \ln \left[ \frac { \lambda^{CP}_{n} ( R , T_{b} )
\: \lambda^{CP'}_{m} ( R , T_{a} ) } { \lambda^{CP}_{n} ( R , T_{a} ) \: \lambda^{CP'}_{m} ( R , T_{b} ) } \right]
\vspace*{2mm} \\ = & \displaystyle \Delta E^{CP;CP'}_{nm}(R)
+ \frac{1}{T_{b}-T_{a}} \: \overline{\alpha} \: e^{ - \overline{\delta} \: T_{a} }
\left( 1 - e^{ - \overline{\delta} \: ( T_{b} - T_{a} ) } \right) \; ,
\end{array}
\end{equation}
where $\overline{\alpha}=\overline{\alpha}_{nm}^{CP;CP'}(R)$ is a suitable combination of the overlaps
and $\overline{\delta}=\overline{\delta}_{nm}^{CP;CP'}(R)$
corresponds to the energy gap to the closest state in the $CP'$ channel for eigenvalue $m$
(which is equivalent to the gap in the $CP$ channel for eigenvalue $n$ to leading order in $1/R$).
In analogy to the fits for the total energies we can improve and stabilize the fits by including the
analogue of eq.~\refc{eqfit2} for the energy differences in the global fit,
\begin{equation}
\label{eqfitdiff2}
- \frac{1}{T_{b}} \: \ln \left[ \frac {\lambda^{CP}_{n} ( R , T_{b} )}{\lambda^{CP'}_{m} ( R , T_{b} } \right] =
 \Delta E^{CP;CP'}_{nm}(R) - \frac{1}{T_{b}} \: \left( \overline{\gamma} +
\overline{\alpha} \: e^{ - \overline{\delta} \: T_{b} } \right) \: ,
\end{equation}
where $\overline{\gamma}=\overline{\gamma}_{nm}^{CP;CP'}(R)=\beta^{CP}_{n}(R)/\beta^{CP'}_m(R)$.

In the analysis we considered fits including all loops, fits for which we left out the data from loops with the
smallest temporal extents, which was beneficial when for the smallest temporal extents the contaminations from higher
excited states were still sizeable, and, in very few cases, fits where the data from the largest temporal extent
has been excluded. The latter was beneficial when the results from the largest temporal extents showed large
fluctuations, but has only been considered when none of the other fits obeyed the quality criteria of
appendix~\ref{app:exc-fits}. In the following we will only use results for which the removal of the excited states
has been successful, i.e. those energies and differences for which at least one of the aforementioned fits passed
the checks of appendix~\ref{app:exc-fits}. We estimated the systematic uncertainty associated with the removal
of the excited states from the difference of the best fit result -- being the one with data from the maximal number
of temporal extents included for which the fit passed the checks -- with the result from a fit where the data from
the loop with the smallest temporal extent of the best fit has been excluded, whenever the latter fit was considered
to be reliable.

\subsection{Results for the flux tube spectrum}

\begin{table*}[t]
\caption{Parameters of the simulations. $t^{\rm sub}_{s,t}$ are the temporal extents of the
sublattices in the LW algorithm and $N^{\rm sub}_{s,t}$ the number of sublattice updates
(see appendix~\ref{app:error-red}). $N_{\rm skip}$ indicates every which point has been used
in a timeslice for the evaluation of Wilson loops, for the purpose of memory reduction
in the simulations. The parameter set at $\beta=5.0$ is the one already discussed in
Ref.~\cite{Brandt:2010bw}.}
\label{tab:sim-paras}
\centering
\begin{tabular}{c|ccc|cccc|cc}
\hline
$\beta$ & $R/a$ & $T/a$ & lattice
& $t^{\rm sub}_s$ & $N^{\rm sub}_s$
& $t^{\rm sub}_t$ & $N^{\rm sub}_t$
& $N_{\rm skip}$ & \# meas \\
\hline
\hline
5.0 & 4-12 & 4 & $32^3$ & 2 & 16000 & 2 & 1500 & 1 & 3200 \\
\cite{Brandt:2010bw} & & 6 & $36^3$ & & & & 2000 & & 3200 \\
 & & 8 & $40^3$ & & & & 6000 & & 5100 \\
 & & 10 & $40^3$ & & & & 12000 & & 6400 \\
 & & 12 & $48^3$ & & & & 16000 & & 8600 \\
\hline
7.5 & 4-24 & 6 & $96\times64^2$ & 2 & 20000 & 4 & 2000 & 4 & 1700 \\
 & & 8 & $128\times64^2$ & 4 & & & 4000 & & 2300 \\
 & & 10 & $80\times64^2$ & 2 & & & 6000 & & 3600 \\
 & & 12 & $96\times64^2$ & 4 & & & 12000 & & 3600 \\
 & & 14 & $112\times64^2$ & 2 & & & 18000 & & 4500 \\
 & & 16 & $128\times64^2$ & 4 & & & 24000 & & 5400 \\
 & & 18 & $144\times64^2$ & 2 & & & 36000 & & 7700 \\
\hline
10.0 & 4-30 & 8 & $128\times96^2$ & 2 & 20000 & 6 & 2000 & 8 & 1800 \\
 & & 10 & $160\times96^2$ & 4 & & & 4000 & & 2500 \\
 & & 12 & $96\times96^2$ & 6 & & & 6000 & & 2700 \\
 & & 14 & $112\times96^2$ & 2 & & & 8000 & & 3500 \\
 & & 16 & $128\times96^2$ & 4 & & & 12000 & & 2900 \\
 & & 18 & $144\times96^2$ & 6 & & & 16000 & & 2200 \\
 & & 20 & $160\times96^2$ & 2 & & & 24000 & & 1400 \\
 & & 22 & $176\times96^2$ & 4 & & & 30000 & & 800 \\
 & & 24 & $192\times96^2$ & 6 & & & 36000 & & 1200 \\
\hline
\end{tabular}
\end{table*}

We perform simulations of 3d $SU(2)$ gauge theory employing the Wilson plaquette action with the common
mixture of heatbath~\cite{Kennedy:1985nu} and three overrelaxation~\cite{Creutz:1987xi} updates. The
simulation parameters together with the parameters of the multilevel
algorithm are collected in Tab.~\ref{tab:sim-paras}. We set the scale using the Sommer parameter
$r_0$~\cite{Sommer:1993ce}, which has been determined for the present parameters with high accuracy in
Ref.~\cite{Brandt:2017yzw}. The energies include a lattice spacing dependent additive
normalization. We get rid of this normalization by subtracting the normalization constant $V_0$
obtained in Ref.~\cite{Brandt:2017yzw} from fitting the potential to the EST prediction.
Both $r_0$ and $V_0$ are listed together with other EST parameters in Tab.~\ref{tab:scale-fits}.
Note, that the ensembles at $\beta=5.0$ have already been analysed in Ref.~\cite{Brandt:2010bw}.
Here we reanalyse the data using the improved fits described above.

\begin{table}
\caption{Results for the Sommer parameter $r_0$, the string tension $\sigma$,
the normalization constant $V_0$ and the boundary coefficient \btt{}
for different $\beta$ values and in the continuum limit from Ref.~\cite{Brandt:2017yzw}.
For \btt{} the first uncertainty is purely statistical, the following systematic
uncertainties are associated with the unknown higher order correction terms in the EST,
the choice of fitrange in the extraction of \btt{} and, for the continuum value, the
continuum extrapolation.}
\label{tab:scale-fits}
 \centering
 \begin{tabular}{c|c|cc|c}
  \hline
  \hline
  & & & & \\[-2.5mm]
  $\beta$ & $r_0/a$ & $\sqrt{\sigma} r_0$ & $aV_0$ & \btt{} \\
  \hline
  \hline
  5.0  & 3.9472(4)(7) & 1.2321(5) & 0.2148(6) & -0.0179({ }5)(50)(23) \\
  7.5  & 6.2860(4)(3) & 1.2341(3) & 0.1740(2) & -0.0244(11)(25)(16) \\
  10.0 & 8.6021(4)(8) & 1.2350(3) & 0.1449(2) & -0.0251({ }5)(27)(22) \\
  \hline
  continuum &  & 1.2356(3)(1) & & -0.0257(3)(38)(17)(3) \\
  \hline
  \hline
 \end{tabular}
\end{table}

\begin{figure}[t]
 \centering
\includegraphics[]{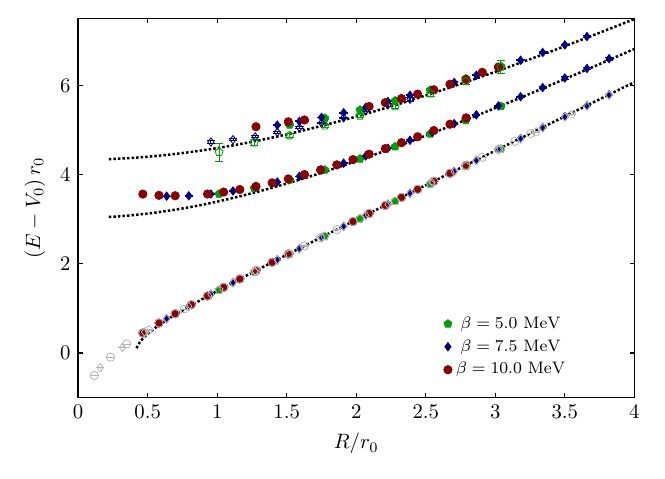}
 \caption{Results for the flux tube spectrum for the lowest three energy levels. To fix the normalization,
 we have subtracted the value of $V_0$ for the individual lattice spacings. We show only results
 for which the contamination from excited states has been removed. The colored open symbols for the
 $n=2$ states are the results for the first excited state in the $(+,+)$-channel. The gray open symbols
 for the groundstate are the results for the potential from Ref.~\cite{Brandt:2017yzw} for comparison.
 The dotted lines are the predictions from the LC spectrum.}
 \label{fig:enes-all}
\end{figure}

We show the results for the energies in Fig.~\ref{fig:enes-all}. Ancillary files
including the jackknife bins of the energies and energy differences are provided along with
this paper. Unfortunately, the removal of contaminations from other states
becomes more and more problematic when going to finer lattices. This is in particular true for
results from excited states in the individual channels, here in particular the first excited
state in the $(+,+)$ channel, so that we could not extract results for this state on our
$\beta=10.0$ ensemble. Where results for three lattice spacings are available, we see that
lattice artifacts are small (see also Figs.~\ref{fig:enes-n1-rsc} and~\ref{fig:enes-n2-rsc}),
so that the continuum results are not expected to differ significantly from the results of the
individual lattice spacings.

\section{Comparison to the EST spectrum}

The energy levels and differences discussed in the previous section can now be compared to the
predictions of the EST.

\subsection{EST predictions for the spectrum}
\label{sec:string-enes}

We first discuss the EST predictions for the energy levels of the flux tube
relevant for this study. For an in-depth discussion of the EST properties
we refer the reader to the reviews~\cite{Aharony:2013ipa,Brandt:2016xsp} and the
extended discussion of section 2 in~\cite{Brandt:2017yzw}.

Using the action up to 6 derivatives order and the constraints for the
coefficients~\cite{Aharony:2009gg,Aharony:2010cx,Billo:2012da,Dubovsky:2012wk}
from Lorentz invariance, the spectrum of a flux tube (open string) of length
$R$ to $O(R^{-5})$ is given by~\cite{Aharony:2010db,Aharony:2011ga}
\be
\label{eq:est-spec}
E^{\rm EST}_{n,l}(R) = E^{\rm LC}_{n}(R) - \bt \frac{\pi^3}{\sqrt{\sigma^3} R^4}
\Big( B_n^l + \frac{d-2}{60} \Big) - \frac{\pi^3 (d-26)}{48 \sigma^2 R^5}
C_n^l + \Ord(R^{-\xi}) \,.
\ee
Following the arguments
from~\cite{Athenodorou:2010cs,Brandt:2010bw,Dubovsky:2012sh,Dubovsky:2014fma},
the first term on the right-hand-side is the full light-cone
spectrum~\cite{Arvis:1983fp}
\be
\label{eq:LC-spectrum}
E^{\rm LC}_{n}(R) = \sigma \: R \: \sqrt{ 1 + \frac{2\pi}{\sigma\:R^{2}} \:
\left( n - \frac{1}{24} \: ( d - 2 ) \right) } \;.
\ee
\btt{} is the dimensionless leading-order
boundary coefficient and $B_n^l$ and $C_n^l$ are dimensionless coefficients
tabulated in table~\ref{tab:3d-string-states} for the lowest few string states.
The $B$ and $C$ coefficients depend on the representation of the state with
respect to rotations around the string axis and lift the degeneracies of the
light-cone spectrum. The lowest order correction term to eq.~\refc{eq:est-spec}
is expected to appear with an exponent $\xi=6$ if the next correction originates
from another boundary term.

\begin{table}
\caption{String states in the lowest four energy levels, their
representation in terms of creation operators $\alpha_m$ in the Fock space
and the values for the coefficients $B_n^l$ and $C_n^l$. Note, that in
three dimensions $C_n^l$ always vanishes.}
\label{tab:3d-string-states}
\centering
\begin{tabular}{cc|lc|cc}
 \hline
 \hline
  & & & & & \\[-2.5mm]
 energy & $\vert n, l \big\rangle$ & representation & $(C,P)$ (in 3d) &
$B_n^l$ & $C_n^l$ \\
 \hline
 \hline
 $E_0$ & $\vert0\big\rangle$ & scalar & $(+,+)$ & 0 & 0
\\
 \hline
 $E_1$ & $\vert1\big\rangle$ & vector & $(+,-)$ & 4 & $d-3$ \\
 \hline
 $E_{2,1}$ & $\vert2,1\big\rangle$ & scalar & $(+,+)$ & 8 & 0 \\
 $E_{2,2}$ & $\vert2,2\big\rangle$ & vector & $(-,-)$ & 32 & $16 (d-3)$ \\
 $E_{2,3}$ & $\vert2,3\big\rangle$ & sym. tracel. tensor & --- & 8 & $4 (d-2)$\\
 \hline
 \hline
\end{tabular}
\end{table}

The effective string theory is expected to break down for $\sqrt{\sigma}R\lesssim 1$,
where the energy of the degrees of freedom reaches the QCD scale.
The EST does not account for several QCD processes, which
are allowed generically in the microscopic theory. Among them are glueball
emission and (virtual) exchange, as well as inner excitations of the flux tube.
The latter are expected to appear as massive excitations on the worldsheet
and are not included in the standard form of the EST. For the static potential,
rigidity or basic massive mode contributions can be included in the EST analysis.
They contaminate the extraction of the boundary coefficient \btt{} and
also add an additional term to the potential (see sections 3 and 5 in
Ref.~\cite{Brandt:2017yzw}). For excited states, the explicit
form of such rigidity or massive mode corrections has not been computed so far,
so that we cannot test the presence of such corrections in this analysis.

\subsection{Comparison of the data to the EST predictions}
\label{sec:string-compare}

To enable the visibility of small differences at large $R$, we from now on plot
rescaled energies and differences following (see also~\cite{Brandt:2009tc})
\begin{equation}
E^{\rm rsc}_{n}(R) = \left( \frac{E_n(R) - V_0}{\sqrt{\sigma}} - \sqrt{\sigma}R \right)
\: \frac{\sqrt{\sigma}R}{\pi} + \frac{1}{24} \,,
\end{equation}
for which the expansion of the LC spectrum, eq.~\refc{eq:LC-spectrum}, to $\Ord(R^{-1})$
yields $E^{\rm rsc}_n=n$ and $\Delta E^{\rm rsc}_{nm}=n-m$.

\begin{figure}[t]
 \centering
\includegraphics[]{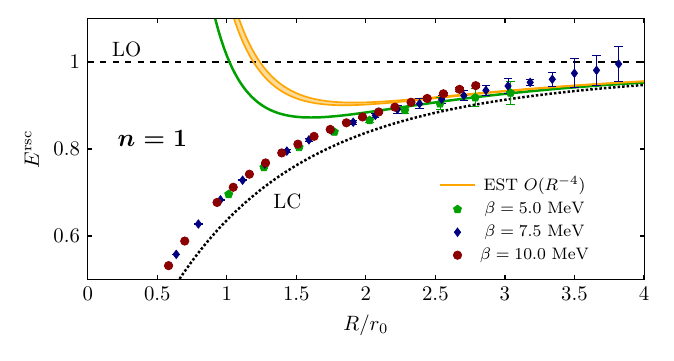}
 \caption{Rescaled results for the energies associated with the first excited string state.
 The yellow band is the EST prediction including the boundary correction with continuum
 parameters, while the green solid line includes the parameters for
 $\beta=5$. The curves with the boundary coefficients for $\beta=7.5$ and 10.0 are very close
 to the continuum curve and therefore have not been included in the plot. The 'LO' curve
 is the EST prediction to $\Ord(R^{-1})$ ($E=n$) and the 'LC' curve the one from the light
 cone spectrum with continuum parameters.}
 \label{fig:enes-n1-rsc}
\end{figure}

\begin{figure}[t]
 \centering
\includegraphics[]{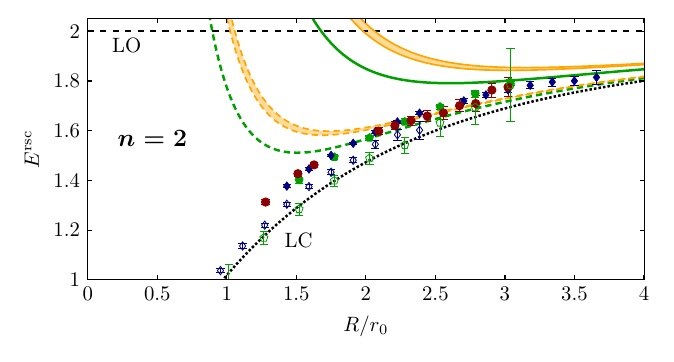}
 \caption{Rescaled results for the energies associated with the second excited string state.
 The curves and points are as in Fig.~\ref{fig:enes-n1-rsc}. The open symbols and dashed lines
 are the results and full EST predictions for the first excited state in the $(+,+)$ channel.}
 \label{fig:enes-n2-rsc}
\end{figure}

\begin{figure}[t]
 \centering
\includegraphics[]{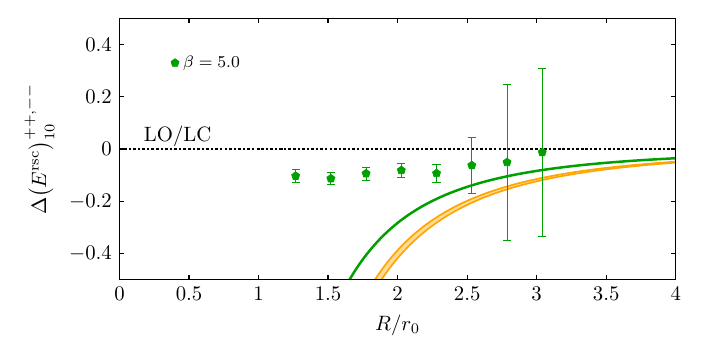}
 \caption{Rescaled energy difference between the first excited state in the $(+,+)$ channel
 and the groundstate in the  $(-,-)$ channel. The dotted line corresponds to LO and LC
 predictions -- which vanish for this difference -- and the curves are the EST prediction
 including the boundary term for $\beta=5.0$ and in the continuum as before.}
 \label{fig:dif-n22-rsc}
\end{figure}

In Figs.~\ref{fig:enes-n1-rsc} and~\ref{fig:enes-n2-rsc} we show the rescaled energies
in comparison to the predictions of the EST including continuum parameters. Note, that differences
in the LC curves for the different lattice spacings would not be visible in the figures. For the
full EST prediction, containing also the boundary term, we also include the curve with the
parameters of $\beta=5.0$. For the other $\beta$-values the full EST curves lie between the
$\beta=5.0$ and continuum curves, but are very close to the latter. Generically, both for
$n=1$ and 2 the data qualitatively follows the
LC curves down to very small values of $R$, where the EST is no longer expected to describe
the flux tube dynamics. Quantitatively, small deviations are visible, which, however, appear
to remain more or less constant with $R$. Note, that this is to some extend an artifact of the
rescaling, eq. (11), which includes subtractions and a multiplication with $R$. Consequently,
a decreasing deviation, like the one in Fig.~\ref{fig:enes-all}, might appear constant in
Figs.~\ref{fig:enes-n1-rsc} to~\ref{fig:dif-n22-rsc}.

For $n=1$, both the $\beta=5.0$ and 7.5 data
become consistent with the full EST prediction around $R\approx2.5 r_0$ (please note the
difference in normalization and the reanalysis for $\beta=5.0$ as compared to the results
presented in Ref.~\cite{Brandt:2018fft}), even though the $\beta=7.5$ data shows a slight
tendency to overshoot the curve for $R\gtrsim 3 r_0$. For $\beta=10.0$
this trend continues and the data lies above the continuum
curve for $R\gtrsim 2.5 r_0$. This might hint to deviations from the curve in the continuum
limit, but it could also be an artifact of insufficient removal of the excited states
contributions which become more severe in this region. In case of the former, it could be a
sign for a higher order corrections, or massive mode contributions, needed to describe the
energies accurately at these distances, eventually
leading to an approach to the $\Ord(R^{-5})$ EST prediction from above. For
$n=2$ the filled points are expected to approach the solid curves, while the open symbols should
approach the dashed ones. This seems to be the case for $\beta=5.0$ and to some extent also
for $\beta=7.5$ and 10.0. For the latter the data from the groundstate in the  $(-,-)$ channel
(filled symbols, corresponding to $\vert2,2\big\rangle$ in Tab.~\ref{tab:3d-string-states})
lie below the EST prediction, but seem to approach it asymptotically. The results for the
first excited state in the $(+,+)$ channel (open symbols, corresponding to
$\vert2,1\big\rangle$ in Tab.~\ref{tab:3d-string-states}) lie below the data from the $(-,-)$
channel, in agreement with the EST predictions, and are much closer to the LC curve.
Note, however, that the agreement of the $\beta=5.0$ with the LC curve
likely is accidental, arising through lattice artifacts, as indicated by the $\beta=7.5$
results. The data approaches the full EST curve for $\beta=5.0$ and $\beta=7.5$, even though
final conclusions are difficult, since reliable results for large $R$-values -- and as such
for $\beta=10.0$ -- are not available for this state.

While conclusions concerning the agreement with the full EST from
the first excited state in the $(+,+)$ channel are difficult, the difference
between this state and the groundstate in the  $(-,-)$ channel might be more sensitive to
the boundary term. Generically ths difference might be useful to investigate the form
of correction terms in the EST as long as they lift the Nambu-Goto degeneracies, since all
universal terms belonging to the $n=2$ states cancel.
Unfortunately, this difference is also very difficult to compute with control over the systematic
effects and so far we have only been able to obtain reliable results at $\beta=5.0$.
Those results, obtained from the individual analysis of the energy differences, are shown
in Fig.~\ref{fig:dif-n22-rsc}. In this normalization the difference appears almost constant with
$R$ while in fact it decreases with $R^{-1}$. It is unclear whether it eventually approaches the
EST prediction for larger $R$ due to the large uncertainties in this region. We note, that the
$R^{-1}$ correction to the LC prediction could well be a sign for a massive mode being responsible
for this difference to be non-vanishing. At the same time it could be a combined contribution of
higher order corrections which mimics such a $R^{-1}$ correction.

\section{Conclusions}

We have extracted the spectrum of the open flux tube in 3d $\SU(2)$ pure gauge theory
up to the second excited state for multiple lattice spacings. The combination of the
multilevel algorithm and a variational method allowed for precise results up to $q\bar{q}$
distances of about $4 r_0$. Excited state contaminations have been removed using a
sophisticated fitting procedure with several checks for systematic effects. The results
qualitatively follow the energy levels of the Nambu-Goto string theory in the light
cone quantization, eq.~\refc{eq:LC-spectrum}, down to small values of $R$ where the
EST is not expected to provide a valid description of the flux tube excitations.
Quantitatively, however, we observe deviations which we compare to the predictions
of the full EST, including a boundary term on top of the Nambu-Goto action with
coefficients computed in Ref.~\cite{Brandt:2017yzw}. We observe that lattice
artifacts are small in general, confirming the findings from Ref.~\cite{Brandt:2009tc}.
However, some lattice artifacts might be visible for the first excited state at large
values of $R$.

While the results tend to agree
with the EST predictions, in particular the results for the second exited state show
the expected splitting predicted by the EST, we observe some deviations for the first
excited state at smaller lattice spacings. This could be a sign for higher order or
massive mode corrections becoming important in the continuum limit, a generic
disagreement with the predictions at large $R$, or uncontrolled systematic effects.
To verify either of these scenarios, further and more accurate results at large $R$
are needed. For $n=2$ the data apparently tends to approach the full EST predictions
asymptotically for all lattice spacings,
even though the approach is slower on the finer lattices. A particularly interesting
quantity with respect to correction terms to Nambu-Goto energy levels in the EST
is the difference between the first excited state in the $(+,+)$ and the groundstate
in the  $(-,-)$ channel, which is vanishing for the LC spectrum. Results with the current
precision -- we were only able to extract results on our coarsest lattice -- decrease
with $R^{-1}$. It is, however, difficult to judge whether the results will first converge
to the boundary correction at large $R$ or quantitatively disagree with this term.

Despite the drastic increase in precision, results for larger values of $R$ and
better precision for the higher excited states are needed to fully confirm or falsify
the agreement between spectrum and EST predictions for the excited states. So far we
could not observe any unambiguous discrepancy between data and EST, despite the fact
that some deviations become apparent on the finer lattices. However, these
could still be remnants of systematic effects, which, generically, become harder to
control for larger values of $R$. Of particular importance for future studies is the
inclusion of possible corrections due to massive modes in the EST predictions,
which in 3d so far have not been computed within the EST framework.

\begin{acknowledgements}
This paper is dedicated to the memory of Pushan Majumdar, a dear colleague and friend.
He introduced me to the world of programming and high-perfomance computing in science,
QCD in particular, and was the co-supervisor of my diploma thesis on QCD strings.
In the years after I visited him several times in India and whenever we met we had lively
discussions on all possible topics connected with physics. QCD strings were one of his
favorite topics and the extension and improvement of our initial study~\cite{Brandt:2009tc},
reported in this article, was always something which was on his mind.

The simulations have
been done in parts on the Athene cluster at the University of
Regensburg and the FUCHS cluster at the Center for Scientific Computing,
University of Frankfurt. I am indebted to the institutes for offering these
facilities. During this work I have received support from DFG via SFB/TRR 55 and
the Emmy Noether Programme EN 1064/2-1.
\end{acknowledgements}

\appendix

\section{Error reduction for Wilson loops}
\label{app:error-red}

The extraction of the flux tube spectrum
requires the accurate computation of large Wilson loops including non-trivial
spatial gluonic operators. For error reduction we use the variant of the L\"uscher-Weisz
multilevel algorithm~\cite{Luscher:2001up} discussed
in~\cite{Brandt:2007iw,Brandt:2009tc} (see also~\cite{Kratochvila:2003zj}).
The multilevel algorithm exploits the locality properties of the Wilson plaquette
action to perform intermediate averages of parts of the operators located in the
interior of sublattices, separated by time-slices with fixed spatial links.
While in the original application of the algorithm to Wilson
loops the spatial operators have been put on the boundaries of the
sublattices~\cite{Luscher:2001up,Majumdar:2002mr}, in the improved algorithm
the spatial operators are located in the middle of a sublattice.

The Wilson loop expectation value can be split into spatial and temporal sublattice
operators, $\mathbb{L}^\alpha_i(x_0)$ and $\mathbb{T}(t)$. The spatial
sublattice operator consists of the spatial operator $S_i^\alpha(\vec{x},x_0)$
from eq.~\refc{eigstates}, located in the middle of the spatial sublattice of extent
$t^{\rm sub}_s$, and two-link operators $\mathcal{T}(x_0)$ of spatial extent $R$ in
direction with unit vector $\vec{i}$,
\begin{equation}
[\mathcal{T}(x_0)]_{abcd} \equiv [U^\ast(\vec{x},x_0)]_{ab} [U(\vec{x}+R\vec{i},x_0)]_{cd} \,,
\end{equation}
connecting the spatial operator with the upper boundary of the sublattice. With
$\mathbb{L}^\dagger$ we denote the operator which includes the spatial operator
$[S_i^\alpha(\vec{x},x_0)]^\dagger$, but connects with the lower boundary of the sublattice
through two-link operators (in an abuse of the notation $\dagger$). The product of two-link
operators is defined by
\begin{equation}
\label{eq:2link-mult}
[\mathcal{T}(x_0)\cdot\mathcal{T}(x_0+a)]_{abcd} = [\mathcal{T}(x_0)]_{aebf}
[\mathcal{T}(x_0+a)]_{ecfd} \,,
\end{equation}
where we have used the sum convention for indices appearing twice on one side of the
equation. The spatial sublattice operator for the spatial sublattice with lower boundary
at time coordinate $x_0$ is then given by
\begin{equation}
[\mathbb{L}^\alpha_i(x_0)]_{ab} = [S_i^\alpha(x_0+t^{\rm sub}_s/2)]_{cd}
[\mathcal{T}(x_0+t^{\rm sub}_s/2)) \cdots \mathcal{T}(x_0+t^{\rm sub}_s-a))]_{cadb} \,.
\end{equation}
The temporal sublattice operator for a temporal sublattice of extent $t^{\rm sub}_t$
with lower boundary at time coordinate $x_0$ consists of the multiplication of
two-link operators from the lower to the upper boundary,
\begin{equation}
[\mathbb{T}(x_0)]_{abcd} \equiv [\mathcal{T}(x_0)\cdot\mathcal{T}(x_0+a)\cdots
\mathcal{T}(x_0+t^{\rm sub}_t-a))]_{abcd} \,.
\end{equation}
Using these two types of sublattice operators and denoting sublattice averages with
$\{\cdot\}$, we can decompose a Wilson loop of
extents $T$ and $R$ as 
\begin{equation}
\langle W_{ij}^\alpha(T,R) \rangle = \langle \{\mathbb{L}^\alpha_i(x_0)\}_{ac}
[\{\mathbb{T}(x_0+t^{\rm sub}_s)\}\cdots \{\mathbb{T}(x_0+T-t^{\rm sub}_t)\}]_{abcd}
\{\mathbb{L}^{\dagger,\alpha}_j(x_0+T)\}_{bc} \rangle .
\end{equation}
Here the temporal extents of the sublattices have to fulfill $T=t^{\rm sub}_s+k\cdot
t^{\rm sub}_t$ with $k\in\mathbb{N}$ and we have made use of the fact that the
temporal sublattice operators obey the two-link operator multiplication law,
eq.~\refc{eq:2link-mult}.

The algorithm contains several parameters that can be tuned to achieve
optimal error reduction. For the sublattices including the temporal sublattice operators
the parameters are the sublattice extent $t^{\rm sub}_t$ and the number of updates
$N^{\rm sub}_t$. Both can be tuned following the lines of Ref.~\cite{Majumdar:2002ga}.
The optimal number of sublattice updates increases with the size of the loops and with
decreasing lattice spacing we found it beneficial to increase $t^{\rm sub}_t$.
For the sublattices including the spatial sublattice operators we similarly
have to tune $t^{\rm sub}_s$ and $N^{\rm sub}_s$. As for the other sublattices we found
it beneficial to increase the former with decreasing lattice spacing. For the excited
states, large values of $N^{\rm sub}_s$ are beneficial, as described in
Ref.~\cite{Brandt:2009tc}. However, when considering 8 different operator sets
including longer contours the computational cost for the computation of the
operators the associate sublattice averaging typically contributes more than 90\% of
the overall computational cost, so that $N^{\rm sub}_s$ cannot become overly large.
To make efficient use of the full temporal extent of the lattice, we vary the temporal
lattice extent for different loops. Since all temporal extents are comparably large, we
do not expect to see relicts of this in the data. Since the algorithm is inherently
memory consuming, we compute the Wilson loops only on a fraction of the points on
a given timeslice for the larger lattices at smaller lattice spacings. When going to
the finer lattices we expect this to not affect uncertainties significantly, since
neighbouring points become more and more correlated.

\nopagebreak

\section{Control of the excited state fits}
\label{app:exc-fits}

\begin{figure}[t]
 \centering
\includegraphics[]{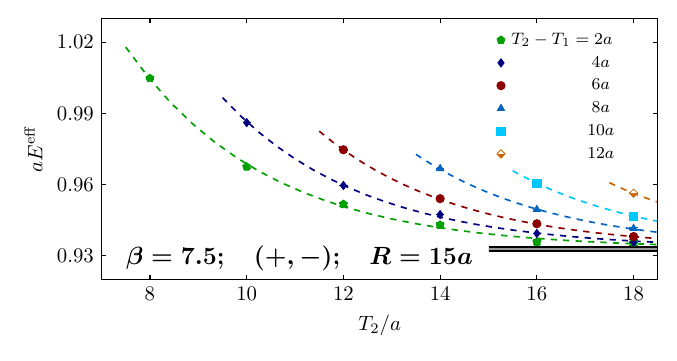}
 \caption{Results for a fit to remove the excited state contributions to the form of
 eq.~\refc{eqfit1} and~\refc{eqfit2} (the latter data and curves are left out to
 make small scale changes visible). The data shown in the plot originates from the
 $\beta=7.5$ ensembles and shows the data in the $(+,-)$-channel at $R=15a$. The black
 band is the final result for the energy.}
 \label{fig:frem-fit}
\end{figure}

The extraction of the spectrum heavily relies on the fits used to remove the contaminations due
to excited states, discussed in Sec.~\ref{sec:exc-state-fits}. A particular example for such a
fit (employing the quality criteria discussed below) is shown in Fig.~\ref{fig:frem-fit}, where
we plot the input effective energies $E^{\rm eff}$ defined by the left-hand-side of
eq.~\refc{eqfit1}, together with the curves obtained from the fit to the central
values. Good control of the systematics of these fits is essential to obtain reliable results.
$\chi^2/{\rm dof}$ typically shows acceptable values even if the fit misses some of the points
at large temporal extents, which have large uncertainties but are the most important ones
concerning the extrapolation. In addition to $\chi^2/{\rm dof}$ we thus install additional
quality criteria and constraints.

We first constrain the fitparameters so that their values will be in the physically relevant
regime. For the fits to extract the energies, eqs.~\refc{eqfit1} and~\refc{eqfit2}, the
relevant parameters are the energies $E$, the overlap ratio $\alpha$, the gap to the first
excited state $\delta$ and the logarithm of the overlap $\gamma$ (here all indices are
suppressed). Both $E$ and $\delta$ should be positive and the latter of the order of the
energy differences between the lowest energy levels. If $\delta$ was an
order of magnitude larger we discarded the fit. The ratio of overlaps $\alpha$ is expected
to be a number of $\Ord(0-100)$ and not much larger. For the extraction of the energy
differences $\Delta E$, eqs.~\refc{eqfitdiff1} and~\refc{eqfitdiff2}, the relevant
parameters are $\overline{\alpha}$, $\overline{\delta}$ and $\overline{\gamma}$. For
$\overline{\alpha}$ and $\overline{\delta}$ similar criteria as for $\alpha$ and $\delta$
apply, whereas the energy difference $\Delta E$ can not be expected to be positive. In most
of the cases they should be, but we also consider differences, in particular the difference
$\Delta E^{++,--}_{10}$, which are expected to be negative.

In addition to these constraints, we also apply the additional quality criteria introduced
in Ref.~\cite{Brandt:2009tc}. In particular, we compare the excited state contribution
from the fit parameters
\begin{equation}
\label{eq:fitcontr1}
\Delta= \frac{1}{T_{b}-T_{a}} \: \alpha_{n} \: e^{ - \delta \: T_{a} } \: 
\left( 1 - e^{ - \delta \: ( T_{b} - T_{a} ) } \right)
\end{equation}
and
\begin{equation}
\label{eq:fitcontr2}
\Delta= - \frac{1}{T_{b}} \: \left( \gamma_n +
\alpha_n \: e^{ - \delta \: T_{b} } \right)
\end{equation}
for eqs.~\refc{eqfit1} and~\refc{eqfit2}, respectively (similar for the energy differences
with $\alpha\to\overline{\alpha}$, $\delta\to\overline{\delta}$ and
$\gamma\to\overline{\gamma}$), to the actual difference of the asymptotic energy with
the effective energy,
\begin{equation}
\label{eqfitcontr3}
\bar{\Delta} = E + \frac{1}{T_{b}-T_{a}} \: \ln \left[ \frac { \lambda ( T_{b} ) } { \lambda ( T_{a} ) } \right]
\end{equation}
and
\begin{equation}
\label{eqfitcontr4}
\overline{\Delta} = E + \frac{1}{T_{b}} \: \ln \left[ \lambda ( T_{b} ) \right]
\end{equation}
for eqs.~\refc{eqfit1} and~\refc{eqfit2}, respectively (for the energy differences $E\to\Delta E$
and the second terms on the r.h.s. are replaced by the terms of the l.h.s. of
eqs.~\refc{eqfitdiff1} and~\refc{eqfitdiff2}). For each fit, we plot the results
for $\Delta$ versus $\overline{\Delta}$ together with the expectation $\Delta=\overline{\Delta}$.
While we allow deviations for the two to three largest values of $\Delta$ and
$\overline{\Delta}$, we only keep those fits for which the other values agree with the
$\Delta=\overline{\Delta}$ line within uncertainties and do not show a systematic trend away from
this line. Example plots for acceptable and non-acceptable fits have been shown in
Ref.~\cite{Brandt:2009tc}.


\end{document}